\documentclass{aa}  
\usepackage{CJK}
\usepackage[colorlinks=true,linkcolor=blue,citecolor=blue, urlcolor=blue]{hyperref}
\usepackage{natbib}
\usepackage{graphicx}
\usepackage{txfonts}
\usepackage{xcolor}
\usepackage[switch]{lineno}

\def\kms{\,\mathrm{km\,s}^{-1}}
\def\ie{{ i.e.,\ }}
\def\eg{{ e.g.,\ }}
\def\lsun{{\rm L}_\odot}
\def\msun{{\rm\,M_\odot}}

\newcommand{\gkai}[1]{\begin{CJK*}{UTF8}{gkai}\raisebox{.1em}{(}#1\raisebox{.1em}{)}\end{CJK*}}
\newcommand{\feh}{\ensuremath{\mathrm{[Fe/H]}}}
\newcommand{\teff}{\ensuremath{T_{\rm eff}} }
\newcommand{\Gaia}{{\emph{Gaia}}}

\makeatletter

\begin{document}

   \title{The Pristine survey -- XXVII. The extremely metal-poor stream C-19 stretches over more than 100 degrees}

    \titlerunning{The C-19 stream over 100$^\circ$}
    \authorrunning{Yuan et al.}

   \author{Zhen Yuan \gkai{袁珍}\inst{1,2,3}, Tadafumi Matsuno \inst{4}, Tatyana Sitnova \inst{5}, Nicolas Martin \inst{3,6},  Rodrigo Ibata \inst{3}, Anke Ardern-Arentsen\inst{7}, Raymond Carlberg\inst{8}, Jonay I. Gonz\'alez Hern\'andez\inst{9,10}, Erika Holmbeck\inst{11}, Georges Kordopatis\inst{12}, Fangzhou Jiang\inst{13,11}, Khyati Malhan\inst{14}, Julio Navarro\inst{15}, Federico Sestito\inst{16}, Kim A. Venn\inst{15}, Akshara Viswanathan\inst{15,17}, Sara Vitali\inst{18}}	   
   
    \institute{School of Astronomy and Space Science, Nanjing University, Nanjing, Jiangsu 210093, China, \texttt{zhen.yuan@nju.edu.cn}
    \and
    Key Laboratory of Modern Astronomy and Astrophysics, Nanjing University, Ministry of Education, Nanjing 210093, China
    \and
    Universit\'e de Strasbourg, CNRS, Observatoire astronomique de Strasbourg, UMR 7550, F-67000 Strasbourg, France
    \and
    Astronomisches Rechen-Institut, Zentrum f\"{u}r Astronomie der Universit\"{a}t Heidelberg, M\"{o}nchhofstras{\ss}e 12-14, 69120 Heidelberg, Germany
    \and
    Institute of Astronomy, Russian Academy of Sciences, Pyatnitskaya 48, 119017, Moscow, Russia
    \and
    Max-Planck-Institut f\"{u}r Astronomie, K\"{o}nigstuhl 17, D-69117 Heidelberg, Germany
    \and
    Institute of Astronomy, University of Cambridge, Madingley Road, Cambridge CB3 0HA, UK
    \and
    Department of Astronomy \& Astrophysics, University of Toronto, Toronto, ON M5S 3H4, Canada
    \and
    Instituto de Astrof\'isica de Canarias, E-38205 La Laguna, Tenerife, Spain
    \and
    Universidad de La Laguna, Departamento de Astrof\'isica, 38206 La Laguna, Tenerife, Spain 
    \and
    The Observatories of the Carnegie Institution for Science, 813 Santa Barbara Street, Pasadena, CA 91101, USA
    \and
     Universit\'e C\^ote d'Azur, Observatoire de la C\^ote d'Azur, CNRS, Laboratoire Lagrange, Nice, France
    \and
    Kavli Institute for Astronomy and Astrophysics, Peking University, Beijing 100871, China
    \and
    DARK, Niels Bohr Institute, University of Copenhagen, Jagtvej 128, 2200 Copenhagen, Denmark
    \and
    Dept. of Physics and Astronomy, University of Victoria, P.O. Box 3055, STN CSC, Victoria BC V8W 3P6, Canada 
    \and
    Centre for Astrophysics Research, Department of Physics, Astronomy and Mathematics, University of Hertfordshire, Hatfield, AL10 9AB, UK
    \and
    Kapteyn Astronomical Institute, University of Groningen, Landleven 12, 9747 AD Groningen, The Netherlands
    \and
    N\'ucleo de Astronom\'ia, Facultad de Ingenier\'ia y Ciencias Universidad Diego Portales, Ej\'ercito 441, Santiago, Chile
    }

\date{Received 13 Feb; accepted 11 Apr}

    \abstract{The discovery of the most metal-poor stream, C-19, provides us with a fossil record of a stellar structure born very soon after the big bang. In this work, we search for new C-19 members throughout the sky by combining two complementary stream-searching algorithms, \texttt{STREAMFINDER} and \texttt{StarGO}, and utilizing low-metallicity star samples from the Pristine survey, as well as Gaia BP and RP spectrophotometric catalogs. We confirm 12 new members, spread over more than 100$^\circ$, using velocity and metallicity information from a set of spectroscopic follow-up programs that targeted a quasi-complete sample of our bright candidates ($G \lesssim 16.0$). From the updated set of stream members, we confirm that the stream is wide, with a stream width of $\sim200$ pc, and dynamically hot, with a derived velocity dispersion of $10.9^{+2.1}_{-1.5} \kms$. The tension remains between these quantities and a purely baryonic scenario in which the relatively low-mass stream (even updated to a few $10^4\msun$) stems from a globular cluster progenitor, as suggested by its chemical abundances. Some heating mechanism, such as preheating of the cluster in its own dark matter halo or through interactions with halo substructures, appears necessary to explain the tension. The impact of binaries on the measured dispersion also remains unknown. Detailed elemental abundances of more stream members, as well as multi-epoch radial velocities from spectroscopic observations, are therefore crucial to fully understanding the nature and past history of the most metal-poor stream of the Milky Way. }

  \keywords{The Galaxy -- Galaxy: abundances -- Galaxy: kinematics and dynamics -- stars: abundances}

\maketitle

\section{Introduction}
\label{sec:intro}

Old stellar streams are fossil relics from the very early Universe. We can find them orbiting in the nearby halo, which allows us to measure the motions and chemical abundances of each individual member with great precision. Detailed chemodynamical studies of streams open the window to understanding the formation of the first stellar structures, as an alternative approach to direct observations such as the images and spectra of high redshift galaxies \citep[\eg GZH2,][]{zavala2024} using the James Webb Space Telescope (JWST). 

As age is a quantity that is very difficult to directly measure for most stars, the iron abundance of a star, which records the chemical enrichment (or lack thereof) of its birth place, is generally used as a crude proxy for age. Extremely metal-poor (EMP) stars have $\feh<-3.0$, which is less than 1/1000th of the solar abundance. These stars were born in a pristine environment at a time when supernova events occurred sparsely and the interstellar medium was not well mixed \citep[\eg][]{argast2000}, roughly $\sim$1 Gyr after the big bang ($z\sim 5-6$). It is not hard to imagine that streams consisting of EMP stars come from stellar structures formed at the same epoch.

In the Gaia era, stellar stream searches \citep[\eg][]{ibata2021} have now revealed a handful of low-metallicity streams \citep{wan2020,martin2022b}, the most metal-poor of which is the C-19 stream \citep{martin2022a}. Originally identified as an EMP stream from the Pristine photometric metallicity survey, follow-up spectroscopic observations of its brightest stars confirmed the stream stars have [Fe/H] = $-$3.28, with an unresolved metallicity dispersion of less than 0.18 dex at the 95$\%$ confidence level. The latter strongly suggests that the C-19 progenitor was a globular cluster. This is also supported by the relatively large variation in light elements between two members, a distinctive signature of globular clusters \citep[see, \eg][and references within]{gratton2004,bastian2018}. Previous observations have also highlighted a tension between the dynamical properties and its presumed globular cluster progenitor. C-19 has a large velocity dispersion for a cluster ($\sigma_v = 6.2^{+2.0}_{-1.4}\kms$), with an estimated mass of only a few times $10^4\msun$, based on precise velocity measurement of ten members brighter than G = 17.6 \citep{yuan2022b}. This is further consolidated by the measurements of 12 faint subgiant members (G $\approx$ 20) by X-Shooter on ESO 8.2 m Very Large Telescope (VLT) \citep{bonifacio2024}. C-19 is the most metal-poor system we have been able to obtain precise chemodynamical measurements for, and current data show that it is a much more complicated structure than expected. 

Previous studies also revealed a new member star separated by 30$^{\circ}$ from the main body \citep{yuan2022b}, which opens the promise that there could be more as-yet undiscovered C-19 members along the orbit. From the perspective of the formation of the first stellar structures, finding more bright stars in the stream would enable us to obtain their detailed elemental abundances and characterize the nature of the stream progenitor. In this work, we report a thorough search for bright C-19 members ($G \lesssim 16.0$) throughout the sky. This task is now possible thanks to the new Gaia DR3 and the BP and RP (BP/RP) spectra that have been used to derive metallicities over the full sky, combined with the improved search capabilities from the fusion of \texttt{STREAMFINDER} and \texttt{StarGO} \citep{yuan2018, ibata2021, yuan2022a}. Using high-resolution spectra from telescopes in both hemispheres, we are able to confirm their memberships based on both velocity and metallicity.

The identification of candidate members is explained in detail in Sec.~\ref{sec:mem}. The analysis of the metallicities of the observed candidates from combined telescope programs is described in Sec.~\ref{sec:met}. The updated dynamical properties of C-19, including all the newly confirmed members, are derived in Sec.~\ref{sec:prop}.
The study of the chemical signature of a subsample of the stars reported here will be presented in a forthcoming paper (Venn et al., in preparation).

\section{Member identification}{\label{sec:mem}}
\subsection{Before Gaia DR3}{\label{subsec:before}}

We first searched for stars similar to known C-19 stars in dynamical space. This was done by applying \texttt{StarGO} to a \texttt{STREAMFINDER} catalog of stars likely to be in a stream, between 10 to 30 kpc. Contrary to the main Gaia EDR3 \texttt{STREAMFINDER} catalog \citep{ibata2021}, we considered stars with lower probabilities, using an 8-$\sigma$ instead of the 10-$\sigma$ originally used by \citet{ibata2021}. \texttt{StarGO} uses a self-organizing map (SOM) to visualize the clustering of these stars in the ($E$, $L_{\rm z}$, $\theta$, $\phi$) space, where the latter two quantities characterize directions of the angular momentum vector: %
\begin{equation}
    \theta = \arccos(L_z/L), \qquad\qquad
    \phi = \arctan(L_x/L_y).
    \label{eq:angles}
\end{equation}

\begin{figure*}
    \centering
     \includegraphics[width=\linewidth]{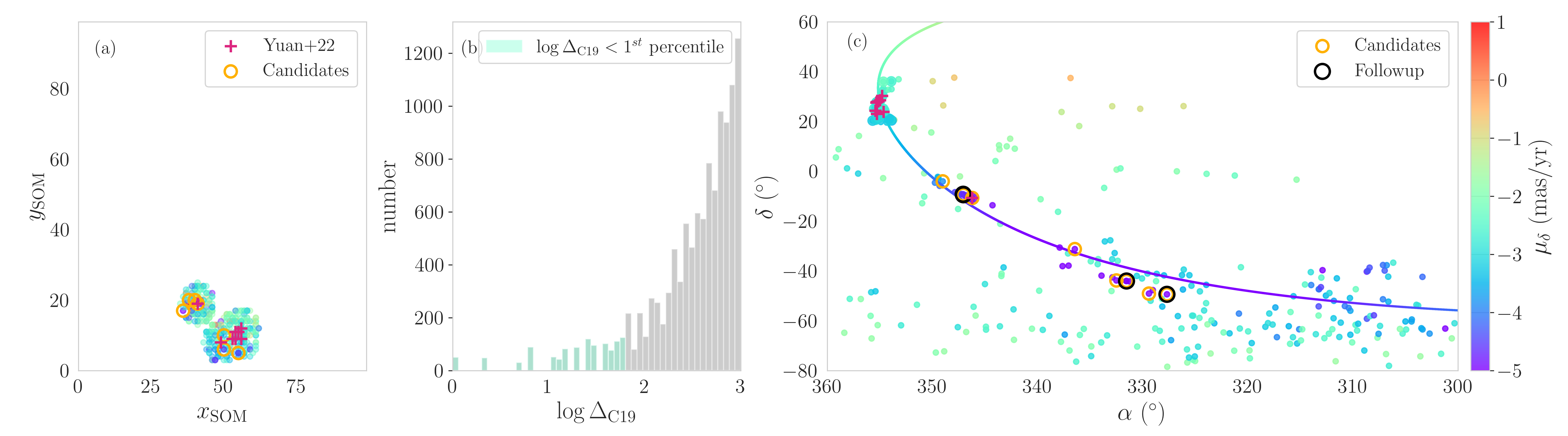}
     \caption{(a) Self-organizing map trained by the \texttt{STREAMFINDER} catalog in the space of ($E$, $L_{\rm z}$, $\theta$, $\phi$). The gray color bar represents the differences in the weight vectors between neighboring neurons. Previously confirmed C-19 members are projected onto the SOM and plotted as red crosses. (b) Histogram of the logarithmic distances in the weight vector space between the existing C-19 stars and stars from the \texttt{STREAMFINDER} catalog ($\Delta_{\rm C19}$). Stars below the first percentile are selected as candidate C-19 members and shown in green in the histogram. They are also plotted as small colored circles in the other panels, with the color representing their proper motions in Gaia DR3 along the declination direction, $\mu_{\delta}$. Candidates brighter than $G = 17.5$ are further highlighted by yellow circles. (c) On-sky projection of the candidate members, color coded by $\mu_{\delta}$. Two stream segments are identifiable with similar $\mu_{\delta}$. The final four candidates for follow-up, selected through other means (see text) are highlighted with black circles.}
    \label{fig:som}
\end{figure*}

The 2D neuron map trained by the \texttt{STREAMFINDER} catalog is shown in Fig.~\ref{fig:som}(a), where a neuron is located at each grid point ($x$, $y$). Stars that have similar properties in dynamical space will be located close to each other on the SOM. This enabled us to easily select stars close to the known C-19 stars on the trained SOM. Instead of remapping these stars using their real measurements (\eg their radial velocities from the follow-up observations), we simply found them in the \texttt{STREAMFINDER} catalog, shown as red crosses in Fig.~\ref{fig:som}(a), so they were more directly comparable to the other stars. In this way, we were not biased by offsets that might arise between the actual velocities and the predicted ones from \texttt{STREAMFINDER}. This procedure is similar to the one we successfully applied to the search for Cetus stream members \citep{yuan2022a}.

To identify new candidate C-19 stars, we first calculated the distance, $\Delta_{\rm C19}$, of each \texttt{STREAMFINDER} star to the ten C-19 members on the 2D map. This is $\Delta_{\rm C19}=\sqrt{(x_{\rm SOM}-x_{\rm SOM, C19})^2+(y_{\rm SOM}-y_{\rm SOM, C19})^2}$, whose distribution is shown in panel (b) of Fig.~\ref{fig:som}. We then selected the 1\% of stars that have the smallest $\Delta_{\rm C19}$. These stars are color coded by their proper motion in the $\delta$ direction in panels (a) and (c) of Fig.~\ref{fig:som}. Their on-sky projection (bottom panel of the Figure) immediately shows two potential stream segments (located at $\delta\sim-10^{\circ}$ and $\delta\sim-45^{\circ}$) that have the same proper motion as that of the stream orbit from \citet{ibata2023}. In total, there are nine stars in these segments brighter than $G = 17.5$, which we highlight with open yellow circles. After further excluding these candidates with photometric metallicities above $\feh=-2.0$ in the SkyMapper DR2 \citep{huang2022} and the Pristine catalog \citep{starkenburg2017}, we are left with four candidates along the C-19 orbit and away from the main body for spectroscopic follow-up. These are highlighted with open black circles in panel (c) of Fig.~\ref{fig:som}. We note that two of these stars are so close together, around $\alpha= 347^{\circ}$, that they are almost indistinguishable in the Figure. We also tested tuning the threshold on $\Delta_{\rm C19}$ to be more inclusive, but this does not lead to any candidates that are more convincing. 

We observed these four most likely candidates using the Magellan telescope on 19 August 2022 (PI: Jiang), with an exposure time of 20 -- 30 minutes for each of the three targets fainter than G = 16.0, and 10 minutes for the most southern one (6559328209695612544) with G = 13.7. We used the standard setup of the Magellan Inamori Kyocera Echelle (MIKE) spectrograph (0.7'' $\times$ 5.0'' slit, 2$\times$2 bin, standard grating angle), which reaches a resolving power of $R$ $\sim$ 28~000 and $\sim$ 35~000 in the red and blue wavelength regions, respectively \citep{bernstein2003}. The data were reduced with the MIKE Carnegie Python pipeline \citep{kelson2003}. We derived their radial velocities using iSpec \citep{blanco2014, blanco2019}, by comparison with standard star HD 182572 observed during the same night. All of the candidates have velocities consistent with the C-19 orbit, listed as stars$^1$ in Tab.\ref{tab:memberRV}. From experience using MIKE, we note that it is very difficult to calibrate velocities extracted from its spectra at better than $\sim$ 2 km s$^{-1}$. We therefore assumed this value as the systematic floor to the MIKE velocity uncertainties, and the statistical uncertainties are denoted as $\delta(v_{\rm r})$ with a star symbol in the upper script in the last column of Tab.\ref{tab:memberRV}. Given the large velocity dispersion of C-19, this choice does impact our inference on the velocity dispersion or the membership. We combined both systematic and statistical uncertainties to derive the velocity dispersion of the stream in Sec.\ref{sec:prop}. These four stars are plotted as orange circles in Fig.~\ref{fig:mem}. Two of them are located close to the member separated by 30$^{\circ}$ away from the main body, which was previously confirmed by \citet{yuan2022b}, and the other two are located 30$^{\circ}$ further south. Their spectra also show that they all have metallicities $\approx-$3.2, which further confirms their membership (see details in Sec.~\ref{sec:met}).

\subsection{After Gaia DR3}{\label{subsec:after}}

\emph{Gaia} DR3 greatly expands the radial velocity sample (RVS) to $G = 16$, with an average magnitude of $G\approx 14.5$ \citep{katz2023}. Moreover, using the Ca H\&K band constructed from the Gaia BP/RP spectra \citep{carrasco2021, andrae2023, montegriffo2023}, we are able to estimate photometric metallicities for very metal-poor stars with $G\lesssim16.0$ \citep[the Pristine-Gaia synthetic catalog;][]{martin2024}. All this new information makes it possible to conduct a star-by-star search for C-19 members, using our knowledge of the orbit and the metallicity of the stream.

We started with stars that have \emph{Gaia} RVS radial velocities along with Pristine-\emph{Gaia} synthetic metallicities $\feh<-2.8$ (using their giant photometric metallicity model). Then we selected those stars along the C-19 trajectory in the 6D kinematic space that have differences in both tangential and radial velocities of less than 30 $\kms$ with respect to the values predicted for the orbit. This approach gives us five potential members with $G < 14.5$. Among them, one is an EMP star discovered in SEGUE \citep{aoki2013} with consistent radial velocity; therefore, it is immediately confirmed. The other candidates are located along the two stream segments previously found, except for one star that is located at $\delta\approx+75^{\circ}$, on the other side of the Milky Way disk. 

We followed up two candidates with Magellan/MIKE in August 2022 (PI: Jiang; the program already mentioned above) and one star with the Isaac Newton Telescope (INT) on 19 July 2022 (PI: Viswanathan). In the INT program, the intermediate dispersion spectrograph (IDS) is equipped with the
RED$+$2 CCD. We used the R1200R grating, a 1.37'' slit width, and the GG495 order-sorting filter. This setting gives a resolving
power of $R$ $\sim$ 8000 around the Ca triplet region from 7850 to 9150 \AA~\citep{viswanathan2025}. The star above the disk was observed with the Subaru telescope on 20 September 2022 (PI: Yuan). In this program, we used the High Dispersion Spectrograph (HDS) in the setup of StdYd, with a slit width of 0.8''. This provides a wavelength coverage of 4000 -- 5340 and 5450 -- 6800 \AA,  $R$ =  45~000 \citep{noguchi2002}. The data were reduced using the IRAF\footnote{IRAF is distributed by the National Optical Astronomy Observatory, which is operated by the Association of Universities for Research in Astronomy (AURA) under a cooperative agreement with the National Science Foundation} script hdsql\footnote{\tiny{http://www.subarutelescope.org/Observing/Instruments/HDS/hdsql-e.html}}. For each target, we took a quick exposure of 10 minutes, irrespective of the telescope. This allowed us to derive their metallicities for further confirmation of their membership, which is described in detail in Sec.~\ref{sec:met}. These five stars are listed as stars$^2$ in Tab.\ref{tab:memberRV}.

Up to now, we are confident that, in addition to the main body of the C-19 stream, there are two additional segments of the stream at lower declination, as revealed in panel (c) of Fig.~\ref{fig:som}. The previously mentioned star-by-star search approach is based on the Gaia RVS sample, which typically contains bright stars with $G < 14.5$. We believe that there should be additional fainter members that are awaiting discovery. Based on the detailed kinematics of the confirmed members in these two stream segments, we were able to do a similar star-by-star search in fainter stars that do not have radial velocity information. To do so, we used both Pristine-\emph{Gaia} synthetic metallicities from Gaia BP/RP and proprietary Pristine metallicities for fainter stars. As before, we selected stars with $\feh<-2.8$ and proper motions compatible with the orbit around the known segments. We find seven candidates with $14.8 <G< 17.4$, only one of which (2641204161744171392) has a radial velocity from Gaia RVS. The velocity agrees well with expectations from the C-19 orbit at this location. 

We observed these seven stars with an exposure time of 10 minutes each on the nights of 2 and 4 September 2023 (PI: Ibata) using the Very Large Telescope (VLT) with the UV-visual echelle (UVES) spectrograph. The slit width was set to 0.7'', which provided a resolving power R $\sim$ 60~000. The wavelength coverage was 3750 -- 5000, 5700 -- 7500, and 7660 -- 9450 \AA\ \citep{dekker2000}. The observed spectra were reduced with the ESO pipeline\footnote{https://www.eso.org/sci/software/pipelines/}. We confirm that all of them have radial velocities consistent with those of the C-19 orbit at their respective positions, listed as stars$^3$ in Tab.\ref{tab:memberRV}. We note that there is one star$^{\divideontimes}$ (2658115921889849472) that has a quite low quality spectrum and has a difference of $\sim$6$\kms$ in radial velocity derived from two bands around 5800 -- 6800\AA\ and 8400 - 8750\AA. The membership is still valid despite the velocity discrepancy. But the spectrum does not have sufficient information to determine its accurate velocity as well as its metallicity. In fact, the metallicity derived from its Ca triplet region in Sec.\ref{sec:met} is above [Fe/H]=$-$3. Therefore, we do not consider this star as a confirmed member and do not take it into account in the analysis of velocity dispersion in Sec.~\ref{sec:prop}.

\begin{figure*}
    \centering
     \includegraphics[width=\linewidth]{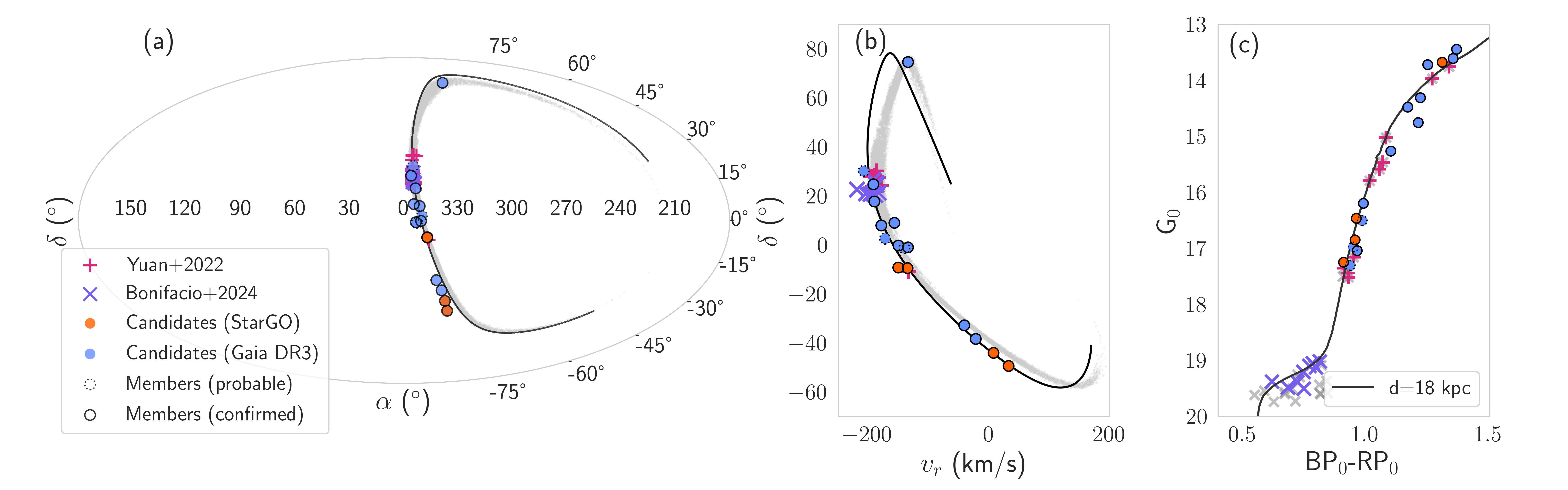}
     \caption{(a) On-sky projection of the updated sample of C-19 stream members. Previously identified members are represented by magenta daggers \citep{yuan2022b} and purple crosses \citep{bonifacio2024}. The solid black line corresponds to the C-19 orbit derived by \citet{ibata2023}, and the small gray dots represent the simulated stream from \citet{errani2022}. Candidates identified from the fusion of \texttt{StarGO} and \texttt{STREAMFINDER} are plotted as orange circles, and the candidates identified using a star-by-star search of photometric metallicity catalogs after \emph{Gaia} DR3 are plotted as blue circles. All shown members have radial velocities consistent with the orbit, as shown in panel (b). Stars that have spectroscopic $\feh<-3.0$ are considered as members and are highlighted with solid black circles. The other stars are probable members and highlighted by dotted circles (see text for more details). (c) Color-magnitude diagram of all the C-19 members previously identified or from this study. The gray crosses denote candidate members without radial velocity information \citep{yuan2022b}. The solid black line here represents the 12-Gyr isochrone with $\feh=-2.2$ from the PARSEC model \citep{bressan2012,fu2018}.}
    \label{fig:mem}
\end{figure*}

\begin{figure}[h]
     \includegraphics[width=\linewidth]{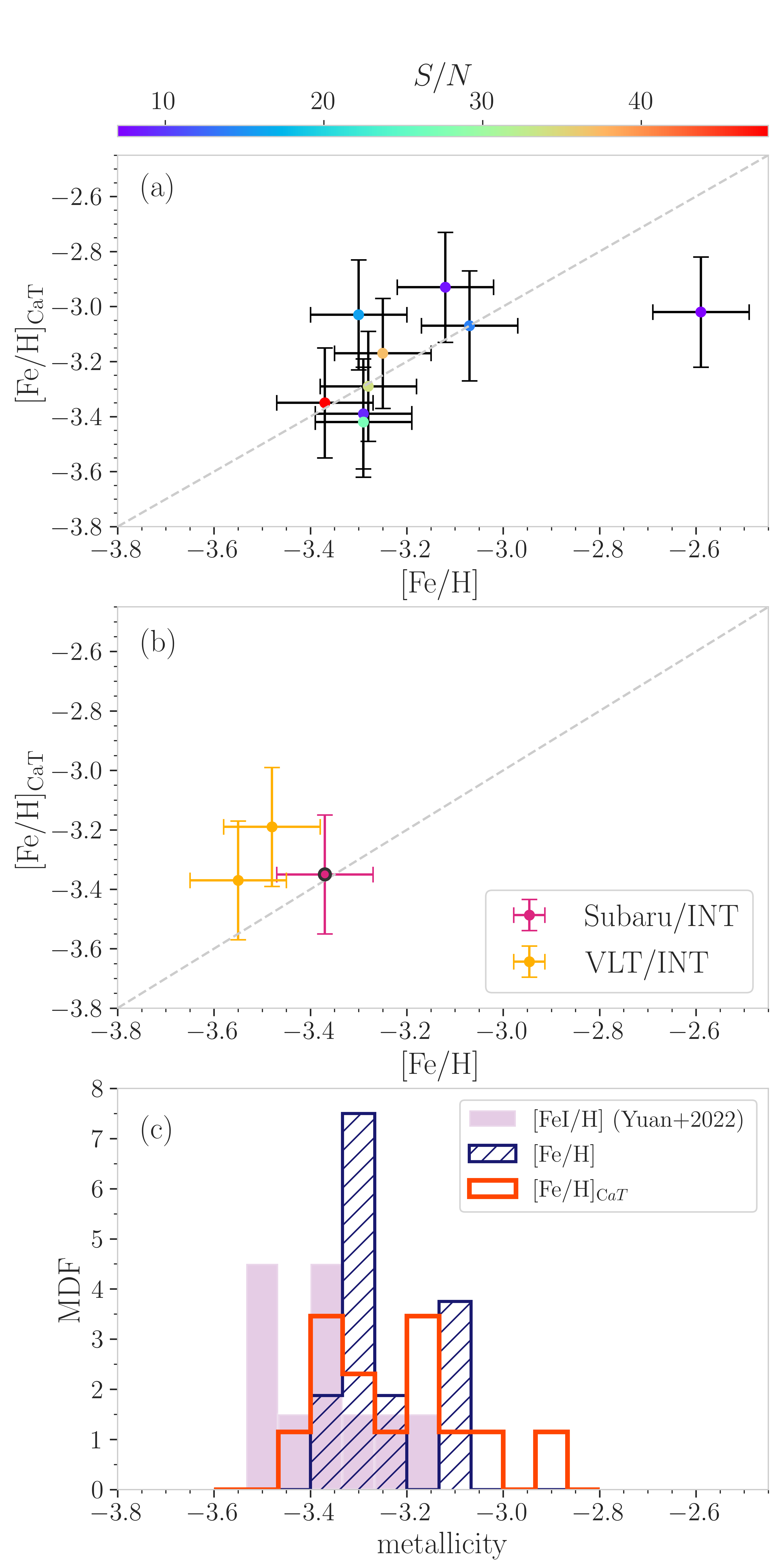}
     \caption{Metallicities of the C-19 members. (a) Comparison of the metallicities derived from the two methods for the same spectra of the nine stars that have high S/N HR spectra. The symbols are color coded by $S/N_1$ (Tab. \ref{tab:memberRV}). (b) Comparison of the metallicities derived from the HR spectra using Fe lines and the low-to-medium spectra using Ca triplet equivalent widths for three C-19 stars. The two existing member stars in \citet{yuan2022b} with VLT/UVES spectra are plotted with yellow symbols and the new member with a Subaru/HDS spectrum is shown in magenta. (c) MDFs of [Fe/H] (blue) and [Fe/H]$_{\rm CaT}$ (orange) in this work compared to the MDF of [FeI/H] of the existing C-19 members from \citet{yuan2022b}.}
    \label{fig:met}
\end{figure}

With all these candidates with confirmed velocities, whose properties are shown in Fig.~\ref{fig:mem}, the C-19 stream is continuously populated from the main body to $\delta = -10^{\circ}$. The other segment starts from $\delta = -38^{\circ}$ and no members are found in between. Overall, the stream profile is more complete above $\delta=-10^{\circ}$, due to the boundaries of the Pristine survey footprint. Within the Pristine footprint, the proprietary Pristine data have more accurate Ca H$\&$K photometry and reach fainter stars compared to the Pristine-Gaia synthetic Ca H$\&$K from Gaia BP/RP. We therefore naturally find more fainter new members ($G \gtrsim 16$) within its sky coverage. The lack of deeper Ca H$\&$K data in the south possibly explains the gap in the current distribution of C-19. The furthest member in the north is separated from the main body of the stream by the Milky Way disk. It is also located in a region without overlap with the Pristine footprint.

\begin{table*}[ht]
\caption{Stellar parameters for the new C-19 member stars}
\begin{center}  
\begin{tabular}{lllllllllll}
\hline
Gaia ID & RA & Dec & G &  Telescope/Instrument& $t_{\rm exp}$ & $S/N_1$\tablefootmark{a} & $S/N_2$\tablefootmark{b} &$v_{\rm r}$ &  $\delta(v_{\rm r})$ \\ 
& ($^\circ$) & ($^\circ$) & & &  (s) & & & ($\mathrm{km\,s^{-1}}$) & ($\mathrm{km\,s^{-1}}$) \\ 
\hline
2820866973064504448$^2$ &353.39287	&17.75368   &13.8	&INT/IDS&600& & 44 &-190.08	&2.44	\\
2758373652717936640$^2$	&354.57313	&9.03539	&14.6	&SDSS&N/A   & & & -156.88	&0.17	\\
2288313499629002624$^2$$^{\dagger}$	&298.74545	&74.65642	&14.1	&Subaru/HDS&600& 48 & 23 &-134.29	&0.67	\\
2606388641446500864$^{1}$	&347.06778	&-9.21263	&17.4	&Magellan/MIKE&1200& 8  & 39 &-150.44	&0.98$^{\ast}$	\\
2606242131522123520$^{1}$	&347.04803	&-9.42817	&17.0 &Magellan/MIKE&1200& 14 & 46 &-135.14	&0.92$^{\ast}$	\\
6600784780223506944$^2$	&339.90245	&-32.79367	&13.5&Magellan/MIKE&600 & 27 & 119 & -41.15	&0.52$^{\ast}$	\\
6559328209695612544$^{1}$	&327.68642	&-49.38001	&13.7&Magellan/MIKE&600 & 37 & 161 &+32.42	&0.54$^{\ast}$	\\
6594796290142997376$^2$	&335.47287	&-38.35307	&14.5	&Magellan/MIKE&600 & 34 & 127 &-22.09	&0.51$^{\ast}$	\\
6567859904530795904$^{1}$	&331.53238	&-44.04167	&16.5&Magellan/MIKE&1800& 16 & 68 &+7.45	&0.83$^{\ast}$	\\
2641204161744171392$^3$	&353.47032	&-0.97001	&14.9	&VLT/UVES&600& 9 & 72 &-134.49	&0.70	\\
2852330082409525760$^3$	&355.83105	&24.77643	&15.3&VLT/UVES&600& 5 & 34 &-191.79	&0.54	\\
2760807387346283648$^3$	&351.31022	&7.96339	&16.4	&VLT/UVES&600& 4 & 36 &-178.60	&0.64	\\
2871048855556252160$^3$	&354.48420	&30.22392	&16.7	&VLT/UVES&600& 2 & 19 &-207.64	&0.61	\\
2658115921889849472$^3$$^{\divideontimes}$	&350.14296	&2.58381	&17.1	&VLT/UVES&600& 3 & 14 &-172.47	&0.83	\\
2644950674600720512$^3$	&350.57152	&-0.22855	&17.1&VLT/UVES&600& 8 & 54 &-150.23	&0.98	\\
2640793013114783872$^3$	&352.99372	&-1.34662	&17.4&VLT/UVES&600& 7 & 44 &-140.23	&0.70	\\
\hline
\end{tabular}
\end{center}
\tablefoot{
The number of the upper script of each star denotes the order of identification in Sec.~\ref{sec:mem}. The star with the dagger symbol has spectra from both Subaru/HDS and INT/IDS, and the radial velocity is derived from the Subaru spectra. The star with the asterisk has low-quality spectra (see details in Sec.~\ref{fig:mem}).
\tablefoottext{a}{S/N per $0.05\,\mathrm{\AA}$ measured at 5863--5865$\AA$.}
\tablefoottext{b}{S/N per $1\,\mathrm{\AA}$ measured at 8639--8644$\AA$.}
For the six stars observed with MIKE, a systematic uncertainty of 2 $\kms$ is assumed and $\delta(v_{\rm r})$ presents statistical uncertainties denoted with an asterisk.
}
\label{tab:memberRV}
\end{table*}

\begin{table*}[ht]
\caption{Metallicities for the new C-19 member stars}
\begin{center}   
\begin{tabular}{lllllllllllllll}
\hline
Gaia ID & T$_{\rm eff}$& $\delta$(T$_{\rm eff}$) & log $g$ & $\delta$(log $g$) & $v_t$ & $M_{K_s}$ & $\delta(M_{K_s})$ & [Fe/H] & [Fe/H]$_{\rm CaT}$ & Member \\ 
 & (K)& (K) & & & ($\mathrm{km\,s^{-1}}$) & & & & & & &\\ 
\hline
2820866973064504448	&4539 &41&0.79&0.10&&-5.10&0.22&& -3.26 &Y\\
2758373652717936640	&   4762 & 56 & 1.33 & 0.13 & & -3.90 & +0.31 && -3.14 & Y \\
2288313499629002624$^{\dagger}$ &	4740&	52	&1.27	&0.12	&+2.31	&-4.03& 0.29 &-3.37	& -3.35 & Y\\
2606388641446500864 &	5479&	110 &3.13	&0.21	&1.31	& +0.42 & 0.51 &-3.12	&-2.93 & Y\\
2606242131522123520 &	5260&	110	&2.59	&0.30	&1.60	& -0.90 & 0.71 &-3.07	&-3.07 & Y\\
6600784780223506944 &	4552&	38	&0.83	&0.09	&2.46	& -5.04 & 0.21 & -3.29	&-3.42 & Y \\
6559328209695612544 &	4668&	51	&1.11	&0.12	&2.40	& -4.40 & 0.28 &-3.25	&-3.17 & Y \\
6594796290142997376 &	4812&	56	&1.45	&0.13	&2.22	& -3.62 & 0.31 & -3.28	&-3.29 & Y \\
6567859904530795904 &	5289&	81	&2.67	&0.22	&1.56	& -0.70 & 0.54 &-3.30	&-3.03 & Y \\
2641204161744171392 &	4832&	76	&1.49	&0.18	&2.20	& -3.50 & 0.43 &-3.29	&-3.39 & Y \\
2852330082409525760 &	4957&	80	&1.74	&0.26	&	    & -2.93 & 0.63 &        &-3.35 & Y \\
2760807387346283648 &	5232&	88	&2.51	&0.24	& 	    & -1.08 & 0.58 &        &-3.20 & Y \\	
2871048855556252160	&   5260&   93  &2.59   & 0.25  &       & -0.89 & 0.61 &        &-2.93 & P\\
2658115921889849472	&   5412&   103 &2.99   & 0.25  &       & +0.07 & 0.60 &        &-2.61 & P\\
2644950674600720512 &	5224&	109	&2.49	&0.30	&	    & -1.12 & 0.71 &        &-3.27 & Y \\
2640793013114783872 &	5417&	111	&2.99	&0.26	&1.39	& +0.08 & 0.62 &-2.59	&-3.02 & P\\
\hline
\end{tabular}
\end{center}
\tablefoot{
The star with the dagger symbol has spectra from both Subaru/HDS and INT/IDS, which give [Fe/H] and [Fe/H]$_{\rm CaT}$ respectively. 
Stars denoted by ''Y'' are confirmed C-19 members that have either [Fe/H] or [Fe/H]$_{\rm CaT}$ within the systematic uncertainty of the C-19 stream (0.2 dex). Those denoted by ''P'' are probable members that have metallicities beyond the systematic uncertainty.}
\label{tab:memberFeH}
\end{table*}

\section{Metallicity}{\label{sec:met}}

For all the 16 stars observed in this work, we estimated the stellar parameters, the distance, and the extinction by fitting the metal-poor ($Z=2\times10^{-5}$, corresponding to $\feh=-3.2$), $\alpha$-enhanced BASTI-IAC isochrones \citep{Hidalgo2018a,Pietrinferni2021a} to the \emph{Gaia} and \emph{2MASS} photometry \citep{cutri2003, skrutskie2006}, as well as the \emph{Gaia} parallax of the star. We maximized the likelihood by varying the distance, initial mass, age, and extinction. We used a flat prior between 10 Gyr and 14 Gyr for the age, and the initial mass function by \citet{Chabrier2003a} for the mass. We added $0.05\,\mathrm{mag}$ error floors in quadrature to the reported uncertainties in photometry, to account for possible systematic offsets between the observations and the isochrone models. The inferred stellar parameters are shown in Table~\ref{tab:memberFeH}.

Combining all the follow-up programs, there are 11 stars that have high-resolution spectra with a resolving power $R\ga$ 30~000 using Subaru/HDS, Magellan/MIKE, and VLT/UVES. Among them, eight stars have sufficiently high signal-to-noise ratio (S/N) spectra for the analysis of individual Fe lines (see Tab.~\ref{tab:memberRV}). The five stars with UVES spectra do not have a high enough S/N for such analysis, but we are still able to infer metallicities from the Ca \textsc{II} triplet (CaT) lines for all of them. In order to check consistency between these two methods, we derived metallicities using the CaT lines for the nine stars with high S/N HR spectra. The two analysis methods are briefly described below.

The first method is based on equivalent widths of individual Fe lines and stellar parameters. We first measured the equivalent widths by fitting a Gaussian profile to each Fe line. We then constructed a model atmosphere by interpolating the grid of MARCS model atmospheres \citep{Gustafsson2008a}, using the stellar parameters of the studied star. Since S/N is not sufficient to determine the microturbulent velocity ($v_t$) for most of our targets, an empirical approach was adopted here. We first fitted a linear function to the relation between $v_t$ and $\log g$ among the extremely metal-poor giants ($\feh<-2.8$, $\teff<5500\,\mathrm{K}$, and $\log g<3.0$) of \citet{Li2022a}. The $v_t$ values we adopted for the C-19 stars were determined from this linear relation with the estimated $\log g$. Based on these inputs ($v_t$, $\teff$, $\log g$, and the equivalent widths), we derived the Fe abundance using the radiative transfer code MOOG \citep{Sneden1973a}, from 7--57 neutral Fe lines. The uncertainties on the derived metallicities are dominated by the uncertainties on \teff and are estimated to be $\sim 0.1\,\mathrm{dex}$.

In the second approach, we first measured the equivalent widths of the CaT lines by fitting a Voigt profile to each component. We then derived the metallicity using the relation between metallicity, total equivalent width, and absolute $K_s$ band magnitude, as calibrated by \citet{Carrera2013a}. The metallicity uncertainties are dominated by the scatter around the calibrated relation ($0.17\,\mathrm{dex}$), yielding a typical uncertainty of $\sim0.2\,\mathrm{dex}$ in metallicity for this method.

The metallicities reported in this work come from these two methods, as well as from spectra with different resolutions. Thus, a consistency check is necessary. We first compared the metallicities of the nine stars with high S/N HR spectra derived using the two methods described above. These stars are color coded by S/N in the top panel of Fig.~\ref{fig:met}. By taking into account the uncertainties, the [Fe/H] metallicities based on the Fe lines and [Fe/H]$_{\rm CaT}$, derived from the CaT lines, are close to the one-to-one line. The most discrepant star (2640793013114783872) corresponds to the lowest S/N spectra ($S/N_1$ = 7 in Tab.~\ref{tab:memberRV}), for which both methods may still work. Excluding this low S/N star, the overall comparison shows that metallicities derived from these two methods are consistent within $\sim0.2$\,dex.

We then compared the metallicities for the same stars that are derived from spectra of different resolutions, as well as from these two methods. There are three stars for which we could do this comparison. In the middle panel of Fig.~\ref{fig:met}, [Fe/H] represents the measurements from Fe lines using the different HR spectra, while [Fe/H]$_{\rm CaT}$ shows estimates using low-to-medium spectra from INT/IDS with $R\sim$ 8~000, covering the CaT region. The different values for these three stars (yellow and magenta symbols) agree well with each other. Here, the two stars with VLT/UVES spectra are existing C-19 members Pristine 355.32+27.59 (2865251577418971392) and Pristine 354.96+28.47 (2866151046649496832), from the C-19 discovery study \citep{martin2022a}. The other star that has one Subaru/HDS spectrum and one INT/IDS spectrum is a new member confirmed in this study. 

\begin{figure}
    \centering
     \includegraphics[width=\linewidth]{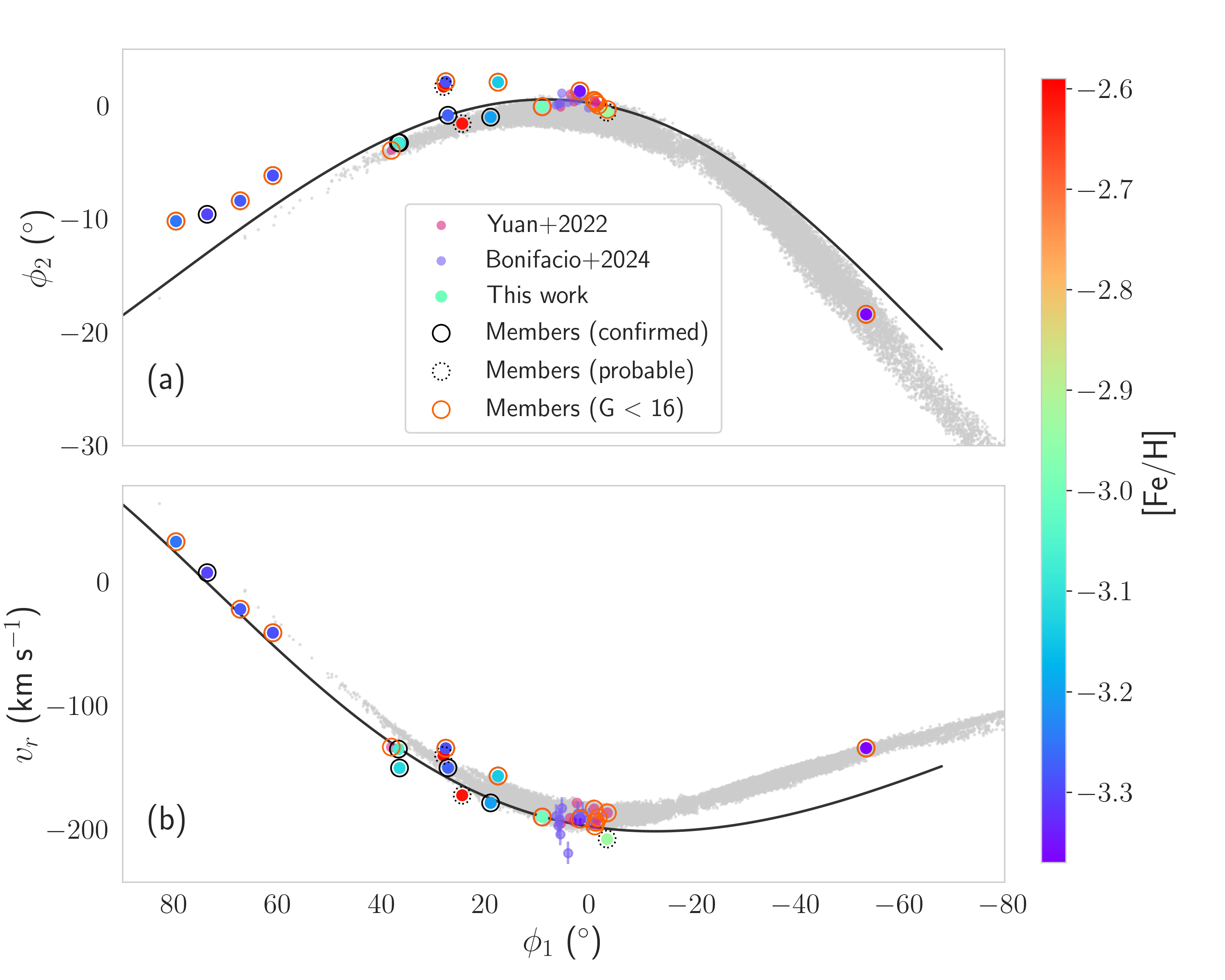}
     \caption{(a) Spatial distribution of the C-19 stream members with radial velocity measurements in the ($\phi_1$, $\phi_2$) coordinate. Similar to Fig.~\ref{fig:mem}, the members from previous studies are plotted as transparent red and purple circles \citep{martin2022b, yuan2022b, bonifacio2024}, along with the derived orbit \citep{ibata2023} and the simulated stream \citep{errani2022}. All the candidates from this work are color coded by their spectroscopic metallicities. The confirmed members have metallicities below $-$3, which are highlighted with solid black circles. The rest are probable members denoted by dotted circles. Among all the confirmed members, those brighter than G $=$ 16 are further highlighted with orange circles, which are just slightly clustered in the main body. (b) The C-19 stream in the ($\phi_1$, $v_{\rm r}$) space, where the radial velocity measurements are taken from \cite{yuan2022b}, \cite{bonifacio2024}, and this work.}
    \label{fig:rot}
\end{figure}

To summarize this section, we selected 17 candidate members of the C-19 stream based on their kinematics in Sec.\ref{sec:mem} and confirm 13 new members that have spectroscopic metallicities consistent within the systematic uncertainty (0.2 dex) of the C-19 stream \citep{martin2022a} when we take into account the uncertainties of our methods. We plot the metallicity distribution (MDF) of [FeI/H] from the C-19 members in \citet{yuan2022b} as the purple histogram, as well as the MDFs of newly confirmed members, in the bottom panel of Fig.~\ref{fig:met}, where the blue and orange histograms denote the MDFs of [Fe/H] and [Fe/H]$_{\rm CaT}$, respectively. The mean of the old members $\langle$[FeI/H]$\rangle$ = $-$3.37, compared to the mean of the new members: $\langle$[Fe/H]$\rangle$ = $-$3.25 and $\langle$[Fe/H]$_{\rm CaT}$$\rangle$ = $-$3.22. The offset is less than the systematic uncertainty of 0.2 dex reported in \citet{martin2022a}. Assuming the metallicity distribution is Gaussian, and using a likelihood analysis similar to that of \citet{martin2022a}, the inference on the metallicity dispersion comes out as consistent with no dispersion for both sets of metallicities listed in Table~2. However, since the spectra of the majority of stars in this paper have rather low S/N, we postpone a detailed discussion on this topic to an upcoming paper (Venn et al., in prep) that reaches the same conclusion, but based on much higher S/N reobservations of several of the bright members presented here. There are three stars with metallicities beyond the systematic uncertainty of the C-19 stream which are not included in the MDFs. These stars still all have low metallicity values ($<-2.5$) and are marked as "probable members" in Tab.~\ref{tab:memberFeH} and highlighted by dotted circles in Fig.~\ref{fig:mem} and Fig.~\ref{fig:rot}. One of them (2658115921889849472$^{\divideontimes}$) has a discrepancy in velocities derived from two bands due to the low-quality spectrum, as mentioned in Sec.~\ref{sec:mem}. We can see that the probable members are not further away from the orbit than the confirmed ones and that they have a similar confidence in membership shown in kinematic space. This may suggest that there was an additional, slightly more metal-rich population in the C-19 progenitor. Intriguingly, this result is similar to those of \citet{bonifacio2024} for turnoff stars in the main body of C-19. However, these stars also all have low S/N spectra and we cannot rule out that the low S/N is unduly affecting our results. Higher S/N spectra are required to confirm their metallicities.

\section{Velocity dispersion, width and mass}{\label{sec:prop}}

\begin{figure}
    \centering
     \includegraphics[width=\linewidth]{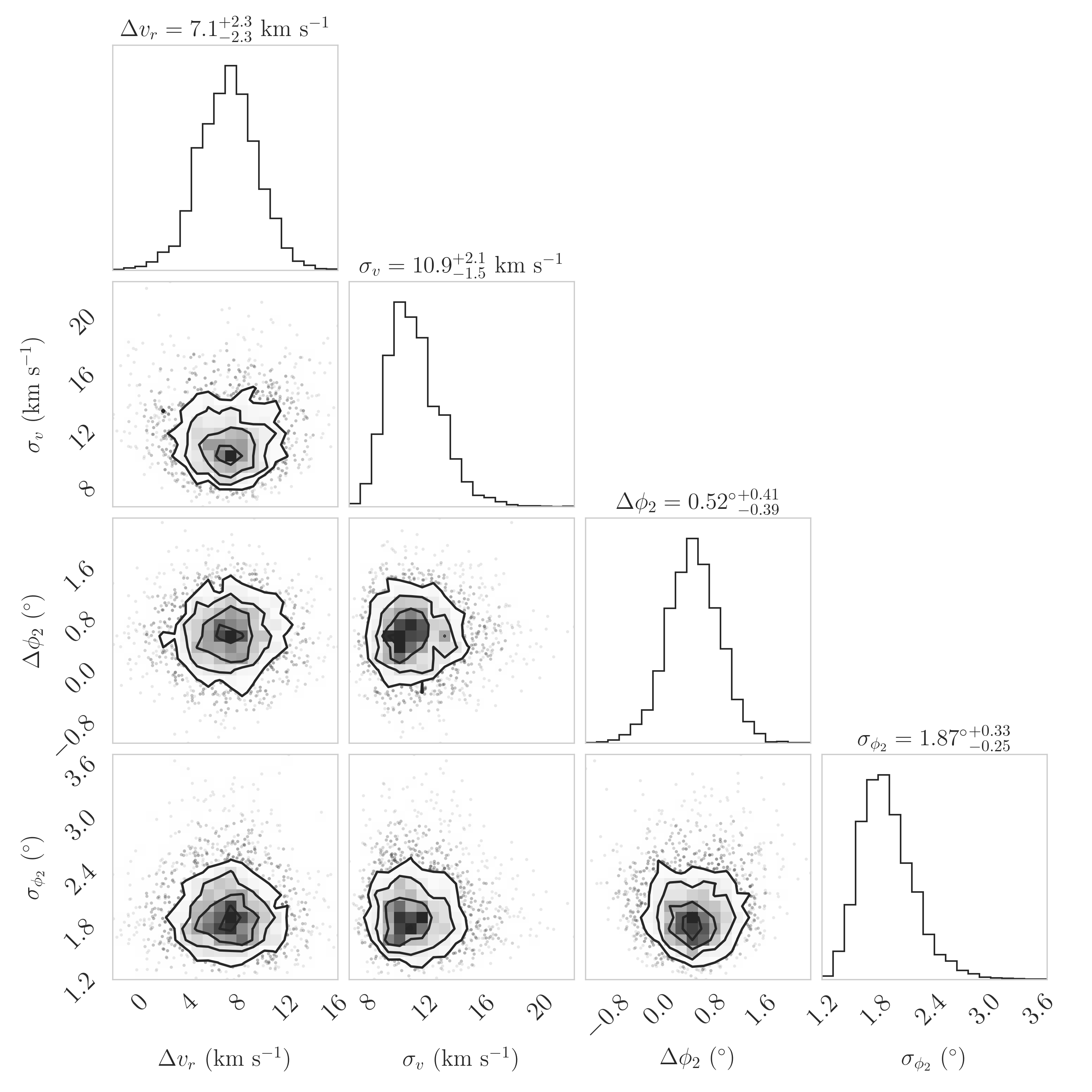}
     \caption{Probability distribution functions (PDFs) of the mean offsets of the 22 confirmed C-19 member stars from the orbit in velocity ($\Delta v$) and position ($\Delta \phi_2$), along with the corresponding dispersions, $\sigma_v$ and $\sigma_{\phi_2}$. The bottom left-hand panels shows the two-dimensional PDF, taken directly from the Markov chain Monte Carlo sampling; the histograms display the marginalized one-dimensional PDFs for the three parameters.}
    \label{fig:corner}
\end{figure}

Based on the full sample of 23 confirmed members with precise radial velocities, 10 stars from \citet{yuan2022b} and 13 stars in Sec.~\ref{sec:met}, we were able to reevaluate the dynamical properties of C-19. Here, we first rotated the equatorial coordinates to a set of stream coordinates ($\phi_1$, $\phi_2$), following the technique presented in \citet{koposov2010} and adopting the pole and zero point of the main body of the stream ($\alpha_0 = 354.356^{\circ}$, $\alpha_{\rm pole} = 81.45^{\circ}$, $\delta_{\rm pole} = -6.346^{\circ}$) from \citet{ibata2023}. The resulting projection of the sample is displayed in Fig.~\ref{fig:rot}, along with that of the stream orbit from the same study, which is simultaneously fitted to 29 streams, for an isolated axisymmetric Milky Way potential \citep{ibata2023}. We then calculated the offsets in velocity ($\Delta v = v_{\rm r} - v_{\rm r, orbit}(\phi_1)$) and perpendicular angular distance ($\Delta \phi_2 = \phi_2 - \phi_{\rm 2, orbit}(\phi_1)$) with respect to the orbit. As in \citet{yuan2022b}, we used the formalism of \citet{martin2018} to infer the mean and dispersion of both quantities, and in this work we obtain $\sigma_{\rm v} = 10.9^{+2.1}_{-1.5}\kms$, shown in Fig.~\ref{fig:corner}. If we simply used a polynomial fitted orbit to the confirmed members, the dispersion is slightly smaller, $\sigma_{\rm v} = 8.5^{+1.5}_{-1.2}\kms$. Among all the members, the star above the disk has the largest offset from the orbit, as shown in the bottom panel of Fig.~\ref{fig:rot}. By excluding this star, the velocity dispersion is only slightly smaller: $\sigma_{\rm v} = 8.9^{+1.6}_{-1.3}\kms$ using the C-19 orbit, $\sigma_{\rm v} = 8.7^{+1.6}_{-1.2}\kms$ using a polynomial fitted orbit. To estimate the stream width, we used the densest part of the stream, \ie the main body located close to the apocenter with the distance around 20 kpc. The dispersion in angular distance is $\sigma_{\phi_2}$ = $0.62^{\circ}$$^{+0.17}_{-0.12}$ and the stream width is translated to be $\sim$ 200 pc. These new results clearly show that, over the 100$^\circ$ traced by the updated sample, the C-19 stream is even hotter than previously found. Previous studies focused on the main body of C-19 and gave a velocity dispersion of $6.2^{+2.0}_{-1.4}\kms$, a stream width of $\sim159$ pc \citep{yuan2022b}, and a velocity dispersion of $5.9^{+3.6}_{-5.9}\kms$ for subgiants \citep{bonifacio2024}. 

The top panel of Fig.~\ref{fig:rot} shows that members brighter than $G=16.0$ (orange circles) are nearly evenly distributed along the orbit and become just slightly more clustered in the main body, where seven bright stars are located. There are now 14 confirmed bright members ($G < 16.0$), which has doubled the previous number of known members presented by \citet{yuan2022b}. There is no reason to expect significant parts of the streams to have been missed at this stage, as stream segments with similar significance and similar orbital properties as C-19 should have been found using our search method in dynamical space by combining \texttt{StarGO} and \texttt{STREAMFINDER}. This is particularly true beyond the current extent of the stream, in the Galactic caps (Fig.~\ref{fig:mem}), where it is easier to isolate streams because of their higher contrast with the Milky Way stellar populations. The one place where parts of the stream could still be evading detection is behind the Milky Way disk, north of the main body of C-19. The upcoming \emph{Gaia} DR4 data release, with deeper BP/RP spectra, along with further extensions of the Pristine footprint in this direction, should help the search for the presence of C-19 in this low-latitude region. 

With the current extension of the stream and the star-by-star search in \Gaia, described in Sec.\ref{subsec:after}, we have checked all the low-metallicity stars in the \texttt{Pristine-Gaia} synthetic catalog along the stream orbit with compatible proper motions down to $G \approx 16.0$ \citep{martin2024}. Assuming the resulting sample of members is close to complete, we can update the estimated mass of the C-19 stream. This mass will still be a lower limit but should be closer to the true mass of the C-19 progenitor than earlier estimates ($>3.5\times10^3\lsun$ or $0.8\times10^4\msun$) based on the main body of the stream \citep{martin2022b}. By summing the fluxes of all the stars brighter than $G = 16.0$ along the giant branch, the total luminosity is $1.8\times10^4\lsun$ after correction for the unobserved fainter stars using the luminosity function associated with the PARSEC isochrone mentioned above \citep{bressan2012,fu2018} and using a Kroupa initial mass function \citep[IMF;][]{kroupa2001, kroupa2002}. Assuming a mass-to-light ratio $M/L =$ 2--3, typical of very old stellar populations \citep{maraston2005}, the mass of C-19 is $\sim$ 3.7 -- 5.5 $\times10^4\msun$, which, unsurprisingly given the much larger number of confirmed bright C-19 members, is significantly larger than the previous estimate. We note that a different IMF would also affect the derived M/L ratio \citep[\eg][]{cappellari2012}. There is observational evidence suggesting that top-heavy IMFs are favored in extremely dense star-forming regions, such as the 30 Doradus star-forming region in the Large Magellanic Cloud \citep{schneider2018} or starburst galaxies at redshift $\sim$2--3 \citep{zhang2018,guo2024}. The C-19 progenitor was very likely a globular cluster, which could have been born in a similar environment of dense clouds and intense star formation. Assuming a top-heavy IMF calibrated for very metal-poor globular clusters \citep{marks2012}, the M/L ratio further increases by a factor of $\sim1.3$, which would result in an even larger progenitor mass of 5 -- 8$\times10^4\msun$.

As discussed in \citet{yuan2022b}, the typical velocity dispersion of a GC stream is less than $5\kms$ \citep{li_S5_2022}. However, with more spectroscopic data available for streams, more have been found to be dynamically hot. For example, the recent velocity dispersion of GD-1 is $\sim$ 7 -- 29 $\kms$ \citep{malhan2019, ibata2023}, despite a progenitor mass of $\sim10^5\msun$ \citep{ibata2020}. It has been suggested that this progenitor was accreted and preprocessed inside a dark matter subhalo \citep{carlberg2020, malhan2021, carlberg2023}. In the case of C-19, the experiment of preprocessing a globular cluster in a dark matter subhalo remains a good explanation for the observed velocity dispersion \citep{errani2022}. A recent study by \citet{carlberg2024} shows the tidal heating of a cold globular cluster stream by $\Lambda$CDM dark subhalos are able to increase the large velocity dispersion to values similar to the dispersion measured in the main body of C-19 ($\sim6\kms$). But the heated velocity dispersion values stay short of the velocity dispersion we find from the larger sample presented in this paper ($\sim10\kms$). Other possible scenarios to explain the high velocity dispersion would include heating by giant molecular clouds \citep[see \eg][]{amorisco16} or, as recently shown from detailed N-body simulations of globular clusters \citep{wang2020, gieles2021, wang2024}, the presence of black holes could enhance the relaxation and disruption of a GC, leading to a larger initial line-of-sight velocity dispersion. Born in a dense and extremely metal-poor region, the C-19 progenitor may also favor a top-heavy IMF, and the strong stellar feedback from overabundant massive stars could cause the cluster to be born dynamically hot in the first place \citep{wang2020,wang2021}. Finally, unresolved binaries could also lead to an artificially large velocity dispersion. From the eight C-19 stars with radial velocity measurements from Gaia, there is no noticeable variation in velocities from our measurements and those from Gaia RVS \citep{katz2023}. We do not see unexpectedly large uncertainties in each measurement either. We note that each star only has one precise velocity measurement from our follow-up and the mean value from Gaia DR3. Therefore, future multi-epoch spectroscopic observations will be crucial for quantifying the contribution of binaries to the observed properties of C-19.

\section{Conclusions}{\label{sec:con}}

In this work, we searched for potential members of C-19 in dynamical space by applying \texttt{StarGO} to the low-significance stream stars from \texttt{STREAMFINDER}. With this technique, we found four new C-19 members, located $30^{\circ}$ and $60^{\circ}$ south from the main body. After Gaia DR3, we performed star-by-star searches along the C-19 orbit using very low-metallicity star catalogs from the Pristine survey (both based on genuine Pristine data and on Gaia XP spectra; \citealt{martin2024}). In total, we confirm 12 new members of the C-19 stream through high-resolution spectroscopy. All new members have velocities compatible with predictions from the orbit based on the main body of the stream \citep{martin2022b,ibata2023}, and their metallicities are, within the uncertainty of measurements, in agreement with the extremely metal-poor mean metallicity of the stream.
 
Our search for C-19 members in the bright regime ($G < 16.0$) utilizes the all-sky Gaia data and we therefore expect that the distribution of these bright stars gives a relatively fair representation of the extent of the stream. Its main body, where it was originally detected, is centered around $\delta\sim+26^{\circ}$ \citep{ibata2021,martin2022a}, and remains the densest part; but we now show that it extends continuously down to $\delta\sim-15^{\circ}$ in the south. A separate segment of the stream of four members is located at $\delta\sim-40^{\circ}$. One single member is found on the other side of the Milky Way disk, at $\delta=+75^{\circ}$, in the north. Overall, the C-19 stream is now shown to extend over $\sim100^{\circ}$ on the sky. By summing up the fluxes from bright stars along the giant branch ($G < 16.0$), we revise the lower mass limit of the stream significantly upward to 3.7 -- 5.5 $\times 10^4\msun$. Based on the velocity measurements of all members, accurate at the $\sim1\kms$ level, we revise upward both the velocity dispersion of the stream to 10.9$^{+2.1}_{-1.5}\kms$, and its Gaussian width to $\sim$ 200 pc. The stream is both hotter and more diffuse than previously reported \citep{martin2022b,yuan2022b}. 

From a dynamical point of view, it remains a puzzle as to how to explain the large velocity dispersion with the relatively small mass of the stream progenitor. Our suggested scenario of a preheating and disruption of the C-19 cluster in its own dark matter halo remains a possibility \citep{errani2022}. In addition, tidal heating from the interaction of the stream with dark matter subhalos \citep{carlberg2024} or giant molecular clouds are also possible. The scenario of a top-heavy IMF would result in an overabundance of massive stars whose stellar feedback could lead to a dynamically hotter initial globular cluster immediately after its birth. Finally, the contribution of binary systems to the observed velocity dispersion remains unknown and could significantly inflate it \citep{wang2024}. It is crucial and timely to obtain multiple spectroscopic observations of known C-19 members to hunt for spectroscopic binaries if we are to understand the dynamical nature of this exceptional stream.

\section{Data availability}{\label{sec:data}}
The updated C-19 member list is provided at https://doi.org/10.5281/zenodo.15240247

\begin{acknowledgements}

ZY, NFM, and RAI acknowledge funding from the European Research Council (ERC) under the European Unions Horizon 2020 research and innovation programme (grant agreement No. 834148). AV gratefully acknowledges support from the Canadian Institute for Theoretical Astrophysics (CITA) through a CITA National Fellowship.

This work has made use of data from the European Space Agency (ESA) mission \Gaia\ (\url{https://www.cosmos.esa.int/gaia}), processed by the \Gaia\ Data Processing and Analysis Consortium (DPAC, \url{https://www.cosmos.esa.int/web/gaia/dpac/consortium}). Funding for the DPAC has been provided by national institutions, in particular the institutions participating in the \Gaia\ Multilateral Agreement.

\end{acknowledgements}



\bibliographystyle{aa}
\bibliography{ms}

\begin{thebibliography}{61}
\expandafter\ifx\csname natexlab\endcsname\relax\def\natexlab#1{#1}\fi

\bibitem[{{Amorisco} {et~al.}(2016){Amorisco}, {G{\'o}mez}, {Vegetti}, \&
  {White}}]{amorisco16}
{Amorisco}, N.~C., {G{\'o}mez}, F.~A., {Vegetti}, S., \& {White}, S. D.~M.
  2016, \mnras, 463, L17

\bibitem[{{Andrae} {et~al.}(2023){Andrae}, {Fouesneau}, {Sordo},
  {Bailer-Jones}, \& {Dharmawardena}}]{andrae2023}
{Andrae}, R., {Fouesneau}, M., {Sordo}, R., {Bailer-Jones}, C.~A.~L., \&
  {Dharmawardena}. 2023, \aap, 674, A27

\bibitem[{{Aoki} {et~al.}(2013){Aoki}, {Beers}, {Lee}, {Honda}, {Ito},
  {Takada-Hidai}, {Frebel}, {Suda}, {Fujimoto}, {Carollo}, \&
  {Sivarani}}]{aoki2013}
{Aoki}, W., {Beers}, T.~C., {Lee}, Y.~S., {et~al.} 2013, \aj, 145, 13

\bibitem[{{Argast} {et~al.}(2000){Argast}, {Samland}, {Gerhard}, \&
  {Thielemann}}]{argast2000}
{Argast}, D., {Samland}, M., {Gerhard}, O.~E., \& {Thielemann}, F.~K. 2000,
  \aap, 356, 873

\bibitem[{{Bastian} \& {Lardo}(2018)}]{bastian2018}
{Bastian}, N. \& {Lardo}, C. 2018, \araa, 56, 83

\bibitem[{{Bernstein} {et~al.}(2003){Bernstein}, {Shectman}, {Gunnels},
  {Mochnacki}, \& {Athey}}]{bernstein2003}
{Bernstein}, R., {Shectman}, S.~A., {Gunnels}, S.~M., {Mochnacki}, S., \&
  {Athey}, A.~E. 2003, in Society of Photo-Optical Instrumentation Engineers
  (SPIE) Conference Series, Vol. 4841, Instrument Design and Performance for
  Optical/Infrared Ground-based Telescopes, ed. M.~{Iye} \& A.~F.~M.
  {Moorwood}, 1694--1704

\bibitem[{{Blanco-Cuaresma}(2019)}]{blanco2019}
{Blanco-Cuaresma}, S. 2019, \mnras, 486, 2075

\bibitem[{{Blanco-Cuaresma} {et~al.}(2014){Blanco-Cuaresma}, {Soubiran},
  {Heiter}, \& {Jofr{\'e}}}]{blanco2014}
{Blanco-Cuaresma}, S., {Soubiran}, C., {Heiter}, U., \& {Jofr{\'e}}, P. 2014,
  \aap, 569, A111

\bibitem[{{Bonifacio} {et~al.}(2024){Bonifacio}, {Caffau}, {Fran{\c{c}}ois},
  {Martin}, {Ibata}, {Yuan}, {Kordopatis}, {Gonz{\'a}lez Hern{\'a}ndez},
  {Aguado}, {Thomas}, {Viswanathan}, {Dodd}, {Gran}, {Starkenburg}, {Lardo},
  {Errani}, {Fouesneau}, {Navarro}, {Venn}, \& {Malhan}}]{bonifacio2024}
{Bonifacio}, P., {Caffau}, E., {Fran{\c{c}}ois}, P., {et~al.} 2024, arXiv
  e-prints, arXiv:2412.20776

\bibitem[{{Bressan} {et~al.}(2012){Bressan}, {Marigo}, {Girardi}, {Salasnich},
  {Dal Cero}, {Rubele}, \& {Nanni}}]{bressan2012}
{Bressan}, A., {Marigo}, P., {Girardi}, L., {et~al.} 2012, \mnras, 427, 127

\bibitem[{{Cappellari} {et~al.}(2012){Cappellari}, {McDermid}, {Alatalo},
  {Blitz}, {Bois}, {Bournaud}, {Bureau}, {Crocker}, {Davies}, {Davis}, {de
  Zeeuw}, {Duc}, {Emsellem}, {Khochfar}, {Krajnovi{\'c}}, {Kuntschner},
  {Lablanche}, {Morganti}, {Naab}, {Oosterloo}, {Sarzi}, {Scott}, {Serra},
  {Weijmans}, \& {Young}}]{cappellari2012}
{Cappellari}, M., {McDermid}, R.~M., {Alatalo}, K., {et~al.} 2012, \nat, 484,
  485

\bibitem[{{Carlberg}(2020)}]{carlberg2020}
{Carlberg}, R.~G. 2020, \apj, 889, 107

\bibitem[{{Carlberg} \& {Agler}(2023)}]{carlberg2023}
{Carlberg}, R.~G. \& {Agler}, H. 2023, \apj, 953, 99

\bibitem[{{Carlberg} {et~al.}(2024){Carlberg}, {Ibata}, {Martin},
  {Starkenburg}, {Aguado}, {Malhan}, {Venn}, \& {Venn}}]{carlberg2024}
{Carlberg}, R.~G., {Ibata}, R., {Martin}, N.~F., {et~al.} 2024, arXiv e-prints,
  arXiv:2410.22966

\bibitem[{{Carrasco} {et~al.}(2021){Carrasco}, {Weiler}, {Jordi}, {Fabricius},
  {De Angeli}, {Evans}, {van Leeuwen}, {Riello}, \&
  {Montegriffo}}]{carrasco2021}
{Carrasco}, J.~M., {Weiler}, M., {Jordi}, C., {et~al.} 2021, \aap, 652, A86

\bibitem[{{Carrera} {et~al.}(2013){Carrera}, {Pancino}, {Gallart}, \& {del
  Pino}}]{Carrera2013a}
{Carrera}, R., {Pancino}, E., {Gallart}, C., \& {del Pino}, A. 2013, \mnras,
  434, 1681

\bibitem[{{Chabrier}(2003)}]{Chabrier2003a}
{Chabrier}, G. 2003, \pasp, 115, 763

\bibitem[{{Cutri} {et~al.}(2003){Cutri}, {Skrutskie}, {van Dyk}, \&
  {Beichman}}]{cutri2003}
{Cutri}, R.~M., {Skrutskie}, M.~F., {van Dyk}, S., \& {Beichman}. 2003, {2MASS
  All Sky Catalog of point sources.}

\bibitem[{{Dekker} {et~al.}(2000){Dekker}, {D'Odorico}, {Kaufer}, {Delabre}, \&
  {Kotzlowski}}]{dekker2000}
{Dekker}, H., {D'Odorico}, S., {Kaufer}, A., {Delabre}, B., \& {Kotzlowski}, H.
  2000, in Society of Photo-Optical Instrumentation Engineers (SPIE) Conference
  Series, Vol. 4008, Optical and IR Telescope Instrumentation and Detectors,
  ed. M.~{Iye} \& A.~F. {Moorwood}, 534--545

\bibitem[{{Errani} {et~al.}(2022){Errani}, {Navarro}, {Ibata}, {Martin},
  {Yuan}, {Aguado}, {Bonifacio}, {Caffau}, {Gonz{\'a}lez Hern{\'a}ndez},
  {Malhan}, {S{\'a}nchez-Janssen}, {Sestito}, {Starkenburg}, {Thomas}, \&
  {Venn}}]{errani2022}
{Errani}, R., {Navarro}, J.~F., {Ibata}, R., {et~al.} 2022, \mnras, 514, 3532

\bibitem[{{Fu} {et~al.}(2018){Fu}, {Bressan}, {Marigo}, {Girardi},
  {Montalb{\'a}n}, {Chen}, \& {Nanni}}]{fu2018}
{Fu}, X., {Bressan}, A., {Marigo}, P., {et~al.} 2018, \mnras, 476, 496

\bibitem[{{Gieles} {et~al.}(2021){Gieles}, {Erkal}, {Antonini}, {Balbinot}, \&
  {Pe{\~n}arrubia}}]{gieles2021}
{Gieles}, M., {Erkal}, D., {Antonini}, F., {Balbinot}, E., \& {Pe{\~n}arrubia},
  J. 2021, Nature Astronomy, 5, 957

\bibitem[{{Gratton} {et~al.}(2004){Gratton}, {Sneden}, \&
  {Carretta}}]{gratton2004}
{Gratton}, R., {Sneden}, C., \& {Carretta}, E. 2004, \araa, 42, 385

\bibitem[{{Guo} {et~al.}(2024){Guo}, {Zhang}, {Yan}, {Gjergo}, {Man}, {Ivison},
  {Fu}, \& {Shi}}]{guo2024}
{Guo}, Z., {Zhang}, Z.-Y., {Yan}, Z., {et~al.} 2024, \apj, 970, 136

\bibitem[{{Gustafsson} {et~al.}(2008){Gustafsson}, {Edvardsson}, {Eriksson},
  {J{\o}rgensen}, {Nordlund}, \& {Plez}}]{Gustafsson2008a}
{Gustafsson}, B., {Edvardsson}, B., {Eriksson}, K., {et~al.} 2008, \aap, 486,
  951

\bibitem[{{Hidalgo} {et~al.}(2018){Hidalgo}, {Pietrinferni}, {Cassisi},
  {Salaris}, {Mucciarelli}, {Savino}, {Aparicio}, {Silva Aguirre}, \&
  {Verma}}]{Hidalgo2018a}
{Hidalgo}, S.~L., {Pietrinferni}, A., {Cassisi}, S., {et~al.} 2018, \apj, 856,
  125

\bibitem[{{Huang} {et~al.}(2022){Huang}, {Beers}, {Wolf}, {Lee}, {Onken},
  {Yuan}, {Shank}, {Zhang}, {Wang}, {Shi}, \& {Fan}}]{huang2022}
{Huang}, Y., {Beers}, T.~C., {Wolf}, C., {et~al.} 2022, \apj, 925, 164

\bibitem[{{Ibata} {et~al.}(2021){Ibata}, {Malhan}, {Martin}, {Aubert},
  {Famaey}, {Bianchini}, {Monari}, {Siebert}, {Thomas}, {Bellazzini},
  {Bonifacio}, {Caffau}, \& {Renaud}}]{ibata2021}
{Ibata}, R., {Malhan}, K., {Martin}, N., {et~al.} 2021, \apj, 914, 123

\bibitem[{{Ibata} {et~al.}(2023){Ibata}, {Malhan}, {Tenachi},
  {Ardern-Arentsen}, {Bellazzini}, {Bianchini}, {Bonifacio}, {Caffau},
  {Diakogiannis}, {Errani}, {Famaey}, {Ferrone}, {Martin}, {di Matteo},
  {Monari}, {Renaud}, {Starkenburg}, {Thomas}, {Viswanathan}, \&
  {Yuan}}]{ibata2023}
{Ibata}, R., {Malhan}, K., {Tenachi}, W., {et~al.} 2023, arXiv e-prints,
  arXiv:2311.17202

\bibitem[{{Ibata} {et~al.}(2020){Ibata}, {Thomas}, {Famaey}, {Malhan},
  {Martin}, \& {Monari}}]{ibata2020}
{Ibata}, R., {Thomas}, G., {Famaey}, B., {et~al.} 2020, \apj, 891, 161

\bibitem[{{Katz} {et~al.}(2023){Katz}, {Sartoretti}, {Guerrier}, {Panuzzo},
  {Seabroke}, {Th{\'e}venin}, \& {Cropper}}]{katz2023}
{Katz}, D., {Sartoretti}, P., {Guerrier}, A., {et~al.} 2023, \aap, 674, A5

\bibitem[{{Kelson}(2003)}]{kelson2003}
{Kelson}, D.~D. 2003, \pasp, 115, 688

\bibitem[{{Koposov} {et~al.}(2010){Koposov}, {Rix}, \& {Hogg}}]{koposov2010}
{Koposov}, S.~E., {Rix}, H.-W., \& {Hogg}, D.~W. 2010, \apj, 712, 260

\bibitem[{{Kroupa}(2001)}]{kroupa2001}
{Kroupa}, P. 2001, \mnras, 322, 231

\bibitem[{{Kroupa}(2002)}]{kroupa2002}
{Kroupa}, P. 2002, Science, 295, 82

\bibitem[{{Li} {et~al.}(2022{\natexlab{a}}){Li}, {Aoki}, {Matsuno}, {Xing},
  {Suda}, {Tominaga}, {Chen}, {Honda}, {Ishigaki}, {Shi}, {Zhao}, \&
  {Zhao}}]{Li2022a}
{Li}, H., {Aoki}, W., {Matsuno}, T., {et~al.} 2022{\natexlab{a}}, \apj, 931,
  147

\bibitem[{{Li} {et~al.}(2022{\natexlab{b}}){Li}, {Ji}, {Pace}, {Erkal},
  {Koposov}, {Shipp}, {Da Costa}, {Cullinane}, {Kuehn}, {Lewis}, {Mackey},
  {Simpson}, {Zucker}, {Ferguson}, {Martell}, {Bland-Hawthorn}, {Balbinot},
  {Tavangar}, {Drlica-Wagner}, {De Silva}, \& {Simon}}]{li_S5_2022}
{Li}, T.~S., {Ji}, A.~P., {Pace}, A.~B., {et~al.} 2022{\natexlab{b}}, \apj,
  928, 30

\bibitem[{{Malhan} {et~al.}(2019){Malhan}, {Ibata}, {Carlberg}, {Valluri}, \&
  {Freese}}]{malhan2019}
{Malhan}, K., {Ibata}, R.~A., {Carlberg}, R.~G., {Valluri}, M., \& {Freese}, K.
  2019, \apj, 881, 106

\bibitem[{{Malhan} {et~al.}(2021){Malhan}, {Valluri}, \& {Freese}}]{malhan2021}
{Malhan}, K., {Valluri}, M., \& {Freese}, K. 2021, \mnras, 501, 179

\bibitem[{{Maraston}(2005)}]{maraston2005}
{Maraston}, C. 2005, \mnras, 362, 799

\bibitem[{{Marks} {et~al.}(2012){Marks}, {Kroupa}, {Dabringhausen}, \&
  {Pawlowski}}]{marks2012}
{Marks}, M., {Kroupa}, P., {Dabringhausen}, J., \& {Pawlowski}, M.~S. 2012,
  \mnras, 422, 2246

\bibitem[{{Martin} {et~al.}(2018){Martin}, {Collins}, {Longeard}, \&
  {Tollerud}}]{martin2018}
{Martin}, N.~F., {Collins}, M. L.~M., {Longeard}, N., \& {Tollerud}, E. 2018,
  \apjl, 859, L5

\bibitem[{{Martin} {et~al.}(2022{\natexlab{a}}){Martin}, {Ibata},
  {Starkenburg}, {Yuan}, {Malhan}, {Bellazzini}, {Viswanathan}, {Aguado},
  {Arentsen}, {Bonifacio}, {Carlberg}, {Gonz{\'a}lez Hern{\'a}ndez}, {Hill},
  {Jablonka}, {Kordopatis}, {Lardo}, {McConnachie}, {Navarro},
  {S{\'a}nchez-Janssen}, {Sestito}, {Thomas}, {Venn}, {Vitali}, \&
  {Voggel}}]{martin2022b}
{Martin}, N.~F., {Ibata}, R.~A., {Starkenburg}, E., {et~al.}
  2022{\natexlab{a}}, \mnras, 516, 5331

\bibitem[{{Martin} {et~al.}(2024){Martin}, {Starkenburg}, {Yuan}, {Fouesneau},
  {Ardern-Arentsen}, {De Angeli}, {Gran}, {Montelius}, {Rusterucci}, {Andrae},
  {Bellazzini}, {Montegriffo}, {Esselink}, {Zhang}, {Venn}, {Viswanathan},
  {Aguado}, {Battaglia}, {Bayer}, {Bonifacio}, {Caffau}, {C{\^o}t{\'e}},
  {Carlberg}, {Fabbro}, {Fern{\'a}ndez-Alvar}, {Gonz{\'a}lez Hern{\'a}ndez},
  {Gonz{\'a}lez Rivera de La Vernhe}, {Hill}, {Ibata}, {Jablonka},
  {Kordopatis}, {Lardo}, {McConnachie}, {Navarrete}, {Navarro}, {Recio-Blanco},
  {Janssen}, {Sestito}, {Thomas}, {Vitali}, \& {Youakim}}]{martin2024}
{Martin}, N.~F., {Starkenburg}, E., {Yuan}, Z., {et~al.} 2024, \aap, 692, A115

\bibitem[{{Martin} {et~al.}(2022{\natexlab{b}}){Martin}, {Venn}, {Aguado}, \&
  {Starkenburg}}]{martin2022a}
{Martin}, N.~F., {Venn}, K.~A., {Aguado}, D.~S., \& {Starkenburg}, E.
  2022{\natexlab{b}}, \nat, 601, 45

\bibitem[{{Montegriffo} {et~al.}(2023){Montegriffo}, {De Angeli}, {Andrae},
  {Riello}, {Pancino}, {Sanna}, {Bellazzini}, {Evans}, {Carrasco}, {Sordo},
  {Busso}, {Cacciari}, {Jordi}, {van Leeuwen}, {Vallenari}, {Altavilla},
  {Barstow}, {Brown}, {Burgess}, {Castellani}, {Cowell}, {Davidson}, {De
  Luise}, {Delchambre}, {Diener}, {Fabricius}, {Fr{\'e}mat}, {Fouesneau},
  {Gilmore}, {Giuffrida}, {Hambly}, {Harrison}, {Hidalgo}, {Hodgkin},
  {Holland}, {Marinoni}, {Osborne}, {Pagani}, {Palaversa}, {Piersimoni},
  {Pulone}, {Ragaini}, {Rainer}, {Richards}, {Rowell}, {Ruz-Mieres}, {Sarro},
  {Walton}, \& {Yoldas}}]{montegriffo2023}
{Montegriffo}, P., {De Angeli}, F., {Andrae}, R., {et~al.} 2023, \aap, 674, A3

\bibitem[{{Noguchi} {et~al.}(2002){Noguchi}, {Aoki}, {Kawanomoto}, {Ando},
  {Honda}, {Izumiura}, {Kambe}, {Okita}, {Sadakane}, {Sato}, {Tajitsu},
  {Takada-Hidai}, {Tanaka}, {Watanabe}, \& {Yoshida}}]{noguchi2002}
{Noguchi}, K., {Aoki}, W., {Kawanomoto}, S., {et~al.} 2002, \pasj, 54, 855

\bibitem[{{Pietrinferni} {et~al.}(2021){Pietrinferni}, {Hidalgo}, {Cassisi},
  {Salaris}, {Savino}, {Mucciarelli}, {Verma}, {Silva Aguirre}, {Aparicio}, \&
  {Ferguson}}]{Pietrinferni2021a}
{Pietrinferni}, A., {Hidalgo}, S., {Cassisi}, S., {et~al.} 2021, \apj, 908, 102

\bibitem[{{Schneider} {et~al.}(2018){Schneider}, {Sana}, {Evans},
  {Bestenlehner}, {Castro}, {Fossati}, {Gr{\"a}fener}, {Langer},
  {Ram{\'\i}rez-Agudelo}, {Sab{\'\i}n-Sanjuli{\'a}n}, {Sim{\'o}n-D{\'\i}az},
  {Tramper}, {Crowther}, {de Koter}, {de Mink}, {Dufton}, {Garcia}, {Gieles},
  {H{\'e}nault-Brunet}, {Herrero}, {Izzard}, {Kalari}, {Lennon}, {Ma{\'\i}z
  Apell{\'a}niz}, {Markova}, {Najarro}, {Podsiadlowski}, {Puls}, {Taylor}, {van
  Loon}, {Vink}, \& {Norman}}]{schneider2018}
{Schneider}, F.~R.~N., {Sana}, H., {Evans}, C.~J., {et~al.} 2018, Science, 359,
  69

\bibitem[{{Skrutskie} {et~al.}(2006){Skrutskie}, {Cutri}, {Stiening}, \&
  {Weinberg}}]{skrutskie2006}
{Skrutskie}, M.~F., {Cutri}, R.~M., {Stiening}, R., \& {Weinberg}, M.~D. 2006,
  \aj, 131, 1163

\bibitem[{{Sneden}(1973)}]{Sneden1973a}
{Sneden}, C. 1973, \apj, 184, 839

\bibitem[{{Starkenburg} {et~al.}(2017){Starkenburg}, {Martin}, \&
  {Youakim}}]{starkenburg2017}
{Starkenburg}, E., {Martin}, N., \& {Youakim}, K. 2017, \mnras, 471, 2587

\bibitem[{{Wan} {et~al.}(2020){Wan}, {Lewis}, {Li}, {Simpson}, {Martell},
  {Zucker}, {Mould}, {Erkal}, {Pace}, {Mackey}, {Ji}, {Koposov}, {Kuehn},
  {Shipp}, {Balbinot}, {Bland-Hawthorn}, {Casey}, {Da Costa}, {Kafle},
  {Sharma}, \& {De Silva}}]{wan2020}
{Wan}, Z., {Lewis}, G.~F., {Li}, T.~S., {et~al.} 2020, \nat, 583, 768

\bibitem[{{Wang}(2020)}]{wang2020}
{Wang}, L. 2020, \mnras, 491, 2413

\bibitem[{{Wang} {et~al.}(2021){Wang}, {Fujii}, \& {Tanikawa}}]{wang2021}
{Wang}, L., {Fujii}, M.~S., \& {Tanikawa}, A. 2021, \mnras, 504, 5778

\bibitem[{{Wang} {et~al.}(2024){Wang}, {Gieles}, {Baumgardt}, {Li}, {Pang}, \&
  {Tang}}]{wang2024}
{Wang}, L., {Gieles}, M., {Baumgardt}, H., {et~al.} 2024, \mnras, 527, 7495

\bibitem[{{Yuan} {et~al.}(2018){Yuan}, {Chang}, {Banerjee}, {Han}, {Kang}, \&
  {Smith}}]{yuan2018}
{Yuan}, Z., {Chang}, J., {Banerjee}, P., {et~al.} 2018, \apj, 863, 26

\bibitem[{{Yuan} {et~al.}(2022{\natexlab{a}}){Yuan}, {Malhan}, {Sestito},
  {Ibata}, {Martin}, {Chang}, {Li}, {Caffau}, {Bonifacio}, {Bellazzini},
  {Huang}, {Voggel}, {Longeard}, {Arentsen}, {Doliva-Dolinsky}, {Navarro},
  {Famaey}, {Starkenburg}, \& {Aguado}}]{yuan2022a}
{Yuan}, Z., {Malhan}, K., {Sestito}, F., {et~al.} 2022{\natexlab{a}}, \apj,
  930, 103

\bibitem[{{Yuan} {et~al.}(2022{\natexlab{b}}){Yuan}, {Martin}, \&
  {Ibata}}]{yuan2022b}
{Yuan}, Z., {Martin}, N.~F., \& {Ibata}, R.~A. 2022{\natexlab{b}}, \mnras, 514,
  1664

\bibitem[{{Zavala} {et~al.}(2024){Zavala}, {Castellano}, {Akins}, {Bakx},
  {Burgarella}, {Casey}, {Ch{\~A}{\textexclamdown}vez Ortiz}, {Dickinson},
  {Finkelstein}, {Mitsuhashi}, {Nakajima},
  {P{\~A}{\textcopyright}rez-Gonz{\~A}{\textexclamdown}lez}, {Arrabal Haro},
  {Bergamini}, {Buat}, {Backhaus}, {Calabr{\~A}{\texttwosuperior}}, {Cleri},
  {Fern{\~A}{\textexclamdown}ndez-Arenas}, {Fontana}, {Franco}, {Grillo},
  {Giavalisco}, {Grogin}, {Hathi}, {Hirschmann}, {Ikeda}, {Jung}, {Kartaltepe},
  {Koekemoer}, {Larson}, {McKinney}, {Papovich}, {Rosati}, {Saito}, {Santini},
  {Terlevich}, {Terlevich}, {Treu}, \& {Yung}}]{zavala2024}
{Zavala}, J.~A., {Castellano}, M., {Akins}, H.~B., {et~al.} 2024, Nature
  Astronomy [\eprint[arXiv]{2403.10491}]

\bibitem[{{Zhang} {et~al.}(2018){Zhang}, {Romano}, {Ivison}, {Papadopoulos}, \&
  {Matteucci}}]{zhang2018}
{Zhang}, Z.-Y., {Romano}, D., {Ivison}, R.~J., {Papadopoulos}, P.~P., \&
  {Matteucci}, F. 2018, \nat, 558, 260

\end{thebibliography}



\appendix


\end{document}